\documentclass[epj]{svjour}
\usepackage{graphicx}
\usepackage{cite}
\usepackage{amssymb,amsmath}
\usepackage{flushend}
\begin{document}
\title{Systematics of Stopping and Flow in Au+Au Collisions}
%\subtitle{Do you have a subtitle?\\ If so, write it here}
\author{
A.~Andronic,\inst{1}
J.~{\L}ukasik,\inst{1,2}
W.~Reisdorf$\,$\inst{1}
\and W.~Trautmann\inst{1}}
%
%\offprints{}         % Insert a name or remove this line
%
\institute{GSI, D-64291 Darmstadt, Germany 
\and IFJ-PAN, Pl-31342 Krak{\'o}w, Poland}
\date{Received: \today}
%\date{Received: date / Revised version: date}
% The correct dates will be entered by Springer
%
\abstract{
Excitation functions of flow and stopping observables for the
Au+Au system at energies from 40 to 1500 MeV per nucleon are presented.
The systematics were obtained by merging the results of the
INDRA and FOPI experiments, both performed at the GSI facility.
The connection to the nuclear equation of state is discussed.
 % end of PACS codes
} %end of abstract
\maketitle
\section{Introduction}
\label{sec:intro}

The study of collective flow in nucleus-nucleus collisions has been an intense
field of research for the past twenty years \cite{reisdorf97, herrmann99}. At
beam energies below several GeV per nucleon, it is mainly motivated by the 
goal to extract the equation of state (EoS) of nuclear matter from the 
quantitative comparison of measurements with the results of microscopic
transport-model calculations \cite{stocker86, danielewicz00, danielewicz02}.
Considerable progress has been made in this direction in recent years but the
constraints on the EoS obtained so far remain rather broad 
\cite{danielewicz02, flow_chapt}.

The results of flow measurements performed before 1999 have been extensively 
reviewed in refs. \cite{reisdorf97, herrmann99}. In the meanwhile, a variety of
new results has become available regarding the directed \cite{reisdorf04,
lemmon99, htun99, liu00, prendergast00, magestro00a,  crochet00, crochet01,
devismes02, uhlig05, westfall01, andronic01, andronic03, magestro00, cussol02,
lukasik05, abd00, chkhaidze01, simic01, bastid04, bastid05 } and elliptic
\cite{lukasik05, abd00, chkhaidze01, simic01, bastid04, bastid05, pinkenburg99,
andronic01npa, chung02, andronic05, stoicea04, lukasik04} flow. These recent
experiments have expanded the study of flow over a broader range of incident
energies. New results became available on collective motion of produced
particles \cite{crochet00, crochet01, devismes02, uhlig05}. Several studies
have focussed on balance (or transition) energies associated with sign changes
of a flow parameter \cite{pinkenburg99, andronic01npa, chung02, andronic05,
magestro00, cussol02, lukasik05}. High statistics measurements allowed to
explore the transverse momentum dependence of flow \cite{liu00, andronic01,
andronic03, bastid05, andronic05}.

Since flow is generated by pressure gradients, it is clear that its
quantitative study reveals aspects of the EoS. However, by itself, flow is not
sufficient to fix the EoS. We need to know, as a function of beam energy, what
density was achieved in the collision. An optimal condition that matter be
piled up to form a dense medium, is that the two colliding ions be stopped in
the course of the collision, before the system starts to expand.
Information on the stopping can be obtained by studying the rapidity density
distributions of the ejectiles in both the beam direction (the original
direction) and the transverse direction. Recently~\cite{reisdorf04}, the ratio
of the variances of the transverse to the longitudinal rapidities was proposed
as an indicator of the degree of stopping and it was found to correlate with
flow provided the incident energy $E/A$ exceeded $150A$ MeV. While this
flow-stopping correlation is only indirectly connected to a pressure-density
correlation, it represents a potentially interesting constraint for microscopic
simulations tending to extract the EoS from heavy ion data.

The main purpose of this review is to present the excitation functions of flow
(directed and elliptic) and of stopping in $^{197}$Au + $^{197}$Au collisions.
This heavy, symmetric system has been studied with a variety of detectors in
the intermediate energy domain throughout the last two decades: 

\begin{center}

\begin{tabular}{llr}
\hline\noalign{\smallskip}
Experiment & Reference &  E/A (MeV)  \\
\noalign{\smallskip}\hline\noalign{\smallskip}
PLASTIC-BALL &  \protect\cite{doss86,doss87,gutbrod89rep,gutbrod90} & 150-1050  \\
MSU-ALADIN &  \protect\cite{tsang93,hsi94,tsang96} & 100-400  \\
LAND-FOPI &  \protect\cite{leifels93} & 400 \\
FOPI &  \protect\cite{ritman95,andronic01npa,andronic05}  & 90-1500  \\
EOS &  \protect\cite{partlan95} & 250-1150  \\
MULTICS-MINIBALL &  \protect\cite{dagostino95,dagostino96} & 35  \\
MSU-4$\pi$ &  \protect\cite{magestro00} & 25-60  \\
INDRA-ALADIN &  \protect\cite{lukasik02,lukasik05,lefevre04} & 40-150  \\
CHIMERA & \protect\cite{pagano05} &  15  \\
\noalign{\smallskip}\hline
\end{tabular}

\end{center}

\noindent The phase space coverage and the range of observables reported in
these studies vary considerably. All these data sets could be and, in most
cases, were indeed used for flow studies. However, except for the comparative
study between the Plastic Ball and the EOS data on directed flow
\cite{partlan95}, and between the Plastic Ball, the FOPI and the INDRA data on
elliptic flow \cite{andronic01npa, lukasik05}, no detailed comparison has been
made so far, in this energy domain, of the results obtained by different
experimental groups with different detectors. 

In this work we will concentrate on the results obtained with the $4\pi$ FOPI
and INDRA detector systems in experiments performed at the heavy-ion
synchrotron SIS at GSI Darmstadt \cite{andronic01, reisdorf04, lukasik05,
andronic05}. The covered ranges of incident energies were $90A$ MeV to $1.5A$
GeV in the FOPI and $40A$ to $150A$ MeV in the INDRA experiments. By combining
the results obtained with the two detectors, having well adapted designs for
the two different energy regimes, we were able to construct coherent
systematics revealing a remarkable evolution of flow and stopping over a large
range of incident energies. 

The observed agreement in the overlap region will serve as a measure of the
absolute accuracy of the experimental data. We will focus on two aspects in
this context, the systematic errors associated with the unavoidable
deficiencies of the experimental devices and on the systematic errors resulting
from the  analysis methods which are not necessarily independent of the former.
Since the two detectors have different acceptances and the reaction mechanism
evolves in the energy region covered by the two experiments, particular
attention will be given to the problem of impact-parameter selection and to the
corrections for the reaction plane dispersion, which need to be adapted
accordingly. For the latter a new method has been devised and applied to the
INDRA data.

\section{The detectors}
\label{sec:detec}

The INDRA detector is constructed as a set of 17 detection rings  with
azimuthal symmetry around the beam axis.  The most forward ring $(2^{\circ}
\leq \theta_{lab}\leq 3^{\circ})$ consists of  12 Si ($300~\mu m$) --
CsI(Tl) (15~cm long) telescopes. The angular range $3^{\circ}$ to $45^{\circ}$
is covered by 8 rings of 192 telescopes  in total, each with three detection
layers: ionization chambers (5~cm of $C_{3}F_{8}$ at 50~mbar), Si-detectors
($300~\mu m$)  and CsI(Tl) scintillators with lengths decreasing from  13.8~cm
to 9~cm with increasing angle.  The remaining 8 rings, covering the region
$45^{\circ} \leq \theta_{lab}\leq176^{\circ}$, have two detection layers:
ionization chambers (5~cm of $C_{3}F_{8}$ at 30~mbar and CsI(Tl) scintillators
(7.6 to 5~cm).  The total granularity is 336 detection cells  covering 90\% of
the 4$\pi$ solid angle.

In the forward region ($\theta_{lab} \leq 45^{\circ}$), ions with
5$\leq$Z$\leq$80 are identified using the $\Delta$E-E method. Over the whole
angular range, isotope identification is obtained for 1$\leq$Z$\leq$4 using the
technique of pulse-shape discrimination  for the CsI(Tl) signals. A complete
technical description of the detector and of its electronics can be found in
\cite{pouthas95}, details of the calibrations performed for the GSI experiments
are given in \cite{lukasik02, trzcinski03}.

The FOPI detector \cite{gobbi93, ritman95} is comprised of two main components:
the forward Plastic Wall and the Central Drift Chamber, covering regions of
laboratory polar angles of 1.2$^\circ <\theta_{lab}<$~30$^\circ$ and 34$^\circ
<\theta_{lab}<$~145$^\circ$, respectively. The Plastic Wall consists of 764
individual plastic scintillator units. Detected reaction products are
identified according to their atomic number, up to $Z\simeq12$,  using the
measured time-of-flight (ToF) and specific energy loss. Particles detected with
the Central Drift Chamber ($Z\leq3$) are identified according to their mass
($A$) by using the measured magnetic rigidity and specific energy loss. The
3-dimensional tracking profits from a high equivalent detector granularity. At
beam energies of $400A$ MeV and above, the forward drift chamber Helitron can
be employed for mass identification of light fragments ($Z\leq2$) at angles
7$^\circ <\theta_{lab}<$~29$^\circ$.

The FOPI detector has an effective granularity exceeding that of INDRA by about
a factor of 4, a property matched to the increasing multiplicity of charged
particles with rising beam energy\footnote{The $4\pi$-integrated charged
particle multiplicities in central collisions increase from typically about 40
at $40A$ MeV to 95 at $150A$ MeV and  exceed 200 (with one quarter of them
being charged pions) at $1.5A$ GeV.}. Both, INDRA and FOPI detectors are
essentially blind to neutral particles, such as neutrons, $\pi^0$ and
$\gamma$'s.  The higher granularity is, however, not the only feature helping
to cope with higher energies. As the energy of the emitted particles rises, a
level is reached where the principle of stopping the particle in a sensitive
detecting material in order to determine its energy is no longer adequate
because the material depth needed leads to a high probability of nuclear
reactions undermining the energy measurement. To avoid this difficulty, one
switches to time-of-flight and magnetic rigidity (in addition to energy loss)
measurements: the apparatus becomes larger and is no longer under vacuum. Hence
the detection thresholds for the various ejectiles are raised. For the FOPI
detectors this means that, {\em e.g.}, at $90A$~MeV fragments with $Z>6$ cannot
be detected at midrapidity anymore.

\section{Impact parameter}
\label{sec:impact}

In a binary collision of massive `objects', the transfer of energy, momentum,
angular momentum, mass etc. between the two partners will be strongly affected
by the impact parameter $b$.  As a consequence one expects to observe large
event-to-event fluctuations due to impact parameter mixing. To be meaningful, a
comparison of experimental observations among each other or with the
predictions of theoretical simulations has to be performed for well defined and
sufficiently  narrow intervals of impact parameter. Generally in microscopic
physics and, in particular, in nuclear physics, the impact parameter is not
directly measurable but has to be estimated from  {\em global} observables
$g$  characterizing the registered events. Global observables are
determined  using all or a significant fraction of the detected particles.

The basic, so-called geometrical model assumption \cite{cavata90}, underlying
the association of an impact parameter $b$ with an observed value $g$ is
that $g$ changes strictly monotonically with $b$ allowing to postulate 
\begin{equation}
\int_{g}^{\infty} \frac{d\sigma(\bar{g})}{d\bar{g}} d\bar{g} 
=\pi b^2(g)
\; \; \; \mbox{or} \; \; 
\int_{0}^{g} \frac{d\sigma(\bar{g})}{d\bar{g}} d\bar{g} 
=\pi b^2(g)
\end{equation}

\noindent where the left (right) hand equation holds for $g$ decreasing
(rising) with $b$. The distribution $d\sigma(g)/dg$ is determined
experimentally in terms of differential cross sections per unit of $g$ in
a minimum bias class of events, {\em i.e.} where a minimum number of conditions
was  required to trigger data taking.

At intermediate to low incident energies, especially for $E/A<100$ MeV, the
literature abounds with an impressive diversity in the choice of global
observables that have been used in attempts to select either narrowly
constrained impact parameters (keywords `highly exclusive' or `ultracentral') 
or events of special interest (keywords `fully equilibrated', `fully
evaporated', `signals of phase coexistence'). The observables vary from very
simple ones like proton, neutron or total charged-particle multiplicity to more
specific ones as, {\em e.g.}, participant proton multiplicity ($N_{p}$)
\cite{gosset77, doss85}, total (E$_{T}$) \cite{phair92, pak96} or light charged
particle transverse kinetic energy (E$_{\perp}^{12}$) \cite{lukasik97}, ratio
of transverse-to-longitudinal kinetic energy ($Erat$) \cite{kuhn93,
reisdorf97a}, degree of isotropy of momenta ($R$) \cite{strobele83,
larochelle96}, transverse momentum directivity ($D$) \cite{beckmann87, bock87,
alard92, phair93}, longitudinal kinetic-energy fraction (E$_{e}$)
\cite{bertch78, cugnon83}, linear momentum transfer \cite{awes81}, total 
kinetic-energy loss ($TKEL$) \cite{charity91, piantelli02}, average parallel
velocity ($V_{av}$) \cite{peter90a}, midrapidity charge ($Z_{y}$)
\cite{ogilvie89a}, total charge of Z$\geq$2 products ($Z_{bound}$)
\cite{hubele91, schuttauf96}, longitudinal component of the quad\-ru\-pole
moment tensor ($Q_{zz}$) \cite{bauer88}. Even more complex observables are
those obtained from sphericity \cite{hanson75, brandt79}, from the kinetic
energy tensor \cite{gyulassy82, buchwald83} or momentum tensor
\cite{wu79,cugnon82, cugnon83}, the thrust ($T$) \cite{farhi77, kapusta81,
cugnon83}, the deflection angle of the projectile ($\Theta_{defl}$)
\cite{buchwald81}, the flow angle ($\Theta_{flow}$) \cite{stocker86,
frankland01}, the location in a `Wilczy{\'n}ski plot' \cite{charity91,
lecolley96, frankland01}, harmonic moments ($H_{2}$) \cite{fox78, fox79,
frankland01}, or combined global variables ($\rho$) \cite{pawlowski97}. The
most sophisticated methods used for impact-parameter selection are based on,
{\em e.g.}, principal component analysis ($PCA$) \cite{desesquelles95,
desesquelles95a, geraci04} or on neural-network techniques ($NN$) \cite{bass94,
david95, haddad97}. 

There are also more technical event selection schemes involving the 
postulation of `complete' events by demanding that nearly the full system 
charge or the full total linear momentum is accounted for. These latter 
methods are specific for a given apparatus since these observables, strictly
constrained by conservation laws, would not be impact parameter selective when
using a perfect detection system. In this case, a comparison of different
experimental data sets at a high level  of precision is difficult and a
comparison with theoretical approaches must  use apparatus specific filter
software that reproduces the hardware cuts  causing the observed selectivity.
In the present study, aiming towards  joining up the data of two rather
different setups, we will try to avoid using such concepts. We will restrict
ourselves to the use of `simple' global observables such as total charged
particle multiplicity $M_c$ or transverse energy $E_{\perp}$ or its variants
$E_{\perp}^{12}$ (limited to $Z\leq 2$) and $Erat$ which, although it involves
also the longitudinal kinetic energy, is highly correlated to $E_{\perp}$ due
to energy conservation constraints.

%------------------------------------------- #b-au-------------------------
\begin{figure}[!htb]
 \leavevmode
\begin{center}
\includegraphics[width=0.9\columnwidth]{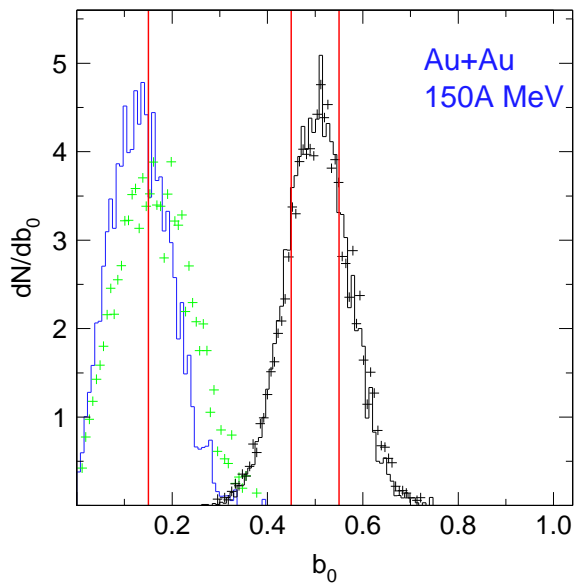}
\end{center}

\caption{Simulated reduced impact parameter distributions for Au+Au collisions
at $150A$ MeV using the global observables $Erat$ (histogram) or charged
particle multiplicity (crosses) for event selection. The two peaks correspond 
to nominal centralities $b_0 < 0.15$ and $0.45 < b_0 < 0.55$, respectively, as
indicated by the vertical lines.}

\label{b-au150}
\end{figure}
%---------------------------------------------------------------------------

The quality of the achieved selectivity in impact parameter is illustrated in
fig.~\ref{b-au150}. It shows distributions of the scaled impact parameter 
$b_0=b/b_{max}$ as obtained from the IQMD transport code \cite{hartnack98} 
simulations for the reaction $^{197}$Au + $^{197}$Au at $150A$ MeV. We take
$b_{max} = 1.15 (A_{P}^{1/3} + A_{T}^{1/3})$~fm and estimate $b$ from the
calculated differential cross sections for the $Erat$ or multiplicity
distributions, using the geometrical sharp-cut approximation. The figure gives
an idea of the achievable impact parameter resolution, typically 1 to 2 fm for
Au on Au, an unavoidable finite size effect. The semi-central event class, at
this energy, happens to be almost invariant against the choice of the selection
method. For the central sample, about 130~mb here, the $Erat$ selection is
somewhat more effective  than the multiplicity selection, an observation
\cite{andronic01npa} found to hold for all higher energies studied with FOPI.
We also conclude that with this selection technique cross section samples
significantly smaller than 100 mb cannot be considered as representative of the
chosen nominal $b$ value.

In this simulation perfect $4\pi$ acceptance was assumed. In reality, 
limitations of the apparatus will further reduce the achievable selectivity.
For the case of FOPI, extensive simulations suggested that the additional loss
of performance is small,  provided the incident energy per nucleon, $E/A$, is
at least 150 MeV  and the considered range of reduced impact parameter does not
significantly exceed $b_0=0.5$.

\begin{figure}[!htb]
% \leavevmode
\begin{center}
\includegraphics[width=0.9\columnwidth]{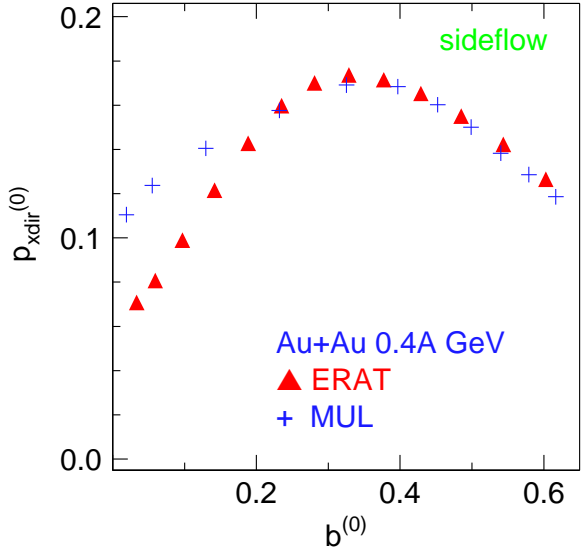}
\end{center}

\caption{Mean charge-integrated ($Z\leq 10$) scaled directed flow,
$p_{xdir}^{(0)}$ measured with FOPI for Au+Au collisions at $400A$ MeV as a
function of the scaled impact parameter $b_0$ as determined with $Erat$
(triangles) and with the multiplicity of charged particles (crosses), after
\protect\cite{reisdorf04}.}

\label{pxdir-prof}
\end{figure}

At sufficiently high $E/A$, the measured directed flow can be used for  a
model-independent comparison of the relative performance of different 
selection methods. This is illustrated in fig.~\ref{pxdir-prof} with  FOPI data
for the reaction Au+Au at $400A$ MeV and for impact-parameter selections using
either $Erat$ or the multiplicity of charged particles in the geometrical
sharp-cut approximation.

%------------------------------------------- #pxdir-prof ------------------
%---------------------------------------------------------------------------

The scaled directed flow is $p_{xdir}^{(0)} \equiv p_{xdir}/u_{1cm}$ where
$p_{xdir}=\sum sign(y) Z u_x/\sum Z$ ($Z$~~fragment charge, $u_{1cm}$ spatial
part of the center of mass projectile 4-velocity, $u_x \equiv \beta_x\gamma$ is
the transverse projection of the fragment 4-velocity on the reaction plane
\cite{danielewicz85}). The sum is taken over all measured charged  particles
with $Z < 10$, excluding pions, and $y$ is the c.m. rapidity. For symmetry
reasons, $p_{xdir}^{(0)}$ has to converge to zero as $b_0 \to 0$. The figure,
therefore, indicates that (i) the $b$ resolution is not perfect in either case
and (ii) for the most central collisions the $Erat$ selection provides a more
stringent impact parameter resolution than the  multiplicity selection, as
already expected on the basis of the simulations (fig.~\ref{b-au150}). The
maximum value of $p_{xdir}^{(0)}$, on the other hand, and the $b_0$ interval
where it is located are robust observables which do not significantly depend on
the selection method. Based on these observations, when FOPI data is analyzed,
in general one employs a mixed multiplicity-$Erat$ strategy for centrality
selection.

Not all global observables behave monotonically with impact parameter, as
evident for $p_{xdir}$ from fig.~\ref{pxdir-prof}. If they are used to select 
central collisions an additional cut is required to suppress  the high $b_0$
branch. Non-monotonic behaviour can also result from losses of heavy ejectiles
close to zero degree or close to target rapidity. These losses tend to increase
with decreasing $E/A$ and (or) increasing $b_0$. In the FOPI case we limit our
analysis to $b_0<0.5$ and require that at least $50\%$ of the total charge has
been identified, a moderate, apparatus specific, constraint that does not
significantly bias the topology of central collisions.

While for particle multiplicities the idea of a monotonic $b$ correlation is
intuitively expected, this is not self-evident for transverse energy. At
sufficiently high energy ($\gtrsim 100A$ MeV), transverse energy is
increasingly generated by the repeated action of many elementary collisions on
the nucleonic level. Since the number of such collisions increases with
increasing target-projectile overlap, high transverse energies are correlated
with {\em low} impact parameters. If $E/A$ is smaller than about 100 MeV mean
field effects involving the system as a whole dominate. One observes
deflections of the projectile-like and target-like remnants to finite polar
angles generating transverse energies that are associated with large impact
parameters which carry large angular momenta. This complication can be avoided
by using  the sum $E_{\perp}^{12}$ of transverse momenta of light charged
particles ($Z \leq 2$) which is more strongly related to the dissipated energy
and does not involve properties of heavier fragments.

%------------------------------------------------------------------------ 

\begin{figure}[!htb]
 \leavevmode
\begin{center}
 \includegraphics[width=0.95\columnwidth]{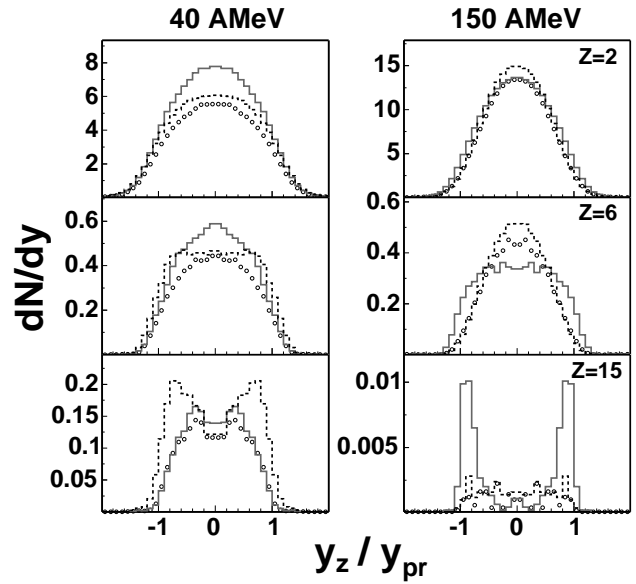}
\end{center}
 
\caption{Longitudinal scaled c.m. rapidity density distributions for $40A$
(left) and $150A$ (right) MeV Au+Au collisions at $b<2$ fm from the INDRA
experiment for selected charges as indicated. Solid histograms - multiplicity
selection of the impact parameter, dashed - $Erat$, circles -
E$_{\perp}^{12}$.}

\label{indra_yz}
 \end{figure}
 
%---------------------------------------------------------------------------

These complexities are illustrated in fig. \ref{indra_yz} using INDRA data for
Au+Au at $40A$ MeV and $150A$ MeV. Shown are charge-separated longitudinal
rapidity distributions for central collisions, selected with three different
observables, multiplicity, $Erat$ and $E_{\perp}^{12}$. To the extent that
stronger yield accumulations near midrapidity  indicate higher centrality, the
multiplicity binning is more  selective of central collisions than $Erat$ at
$40A$ MeV while at $150A$ MeV the reverse is true. This appears more pronounced
for the  cases of larger fragments shown in the lower panels.

For the $Erat$ and $E_{\perp}^{12}$ selections in fig. \ref{indra_yz}  the
centrality has been defined by all the relevant reaction products except the
one of interest. This method of excluding the `particle of interest' (POI) from
the selection criteria allows to avoid autocorrelations between the studied
observable and the one used for the estimation of centrality.  On the other
hand, the exclusion of the POI makes the observable used for the impact
parameter selection particle dependent, {\em i.e.} no longer globally event
dependent.  This may affect the partitions belonging to a given centrality bin
since, depending on the particle, the event may, or may not fulfill the
criteria for a given centrality class. It has serious consequences when the
autocorrelation is strong, especially for low energy collisions which are
characterized by the presence of intermediate and  heavy mass fragments
carrying substantial amounts of momentum.  Excluding, or missing, such a
fragment unavoidably affects the measure  of the impact parameter and increases
its fluctuations. $E_{\perp}^{12}$  does not depend on the exclusion or
detection of heavy fragments and thus is better suited for lower energies.

On the other hand, in the case of the INDRA detector, the multiplicity
observable does not seem to be the optimal centrality selector at high energies
(fig. \ref{indra_yz}, right bottom panel) where due to inefficiencies for light
particles (multi-hits, punch throughs), this observable may admix less central
events with higher multiplicities of fragments to the most central bin.  Using
$E_{\perp}^{12}$ as a centrality selector avoids switching the selection
method  when studying excitation functions. As can be seen in the figure,
$E_{\perp}^{12}$ performs similar to multiplicity at $40A$ MeV and similar to
$Erat$ at $150A$ MeV. Since molecular dynamics simulations confirm this 
observation \cite{plagnol00}, we choose $E_{\perp}^{12}$ in the following as a
centrality measure for the INDRA data, unless indicated otherwise.

\section{Rapidity density and stopping}
\label{sec:rap_dens}

\begin{figure}[!htb]
 \leavevmode
 \begin{center}
 \includegraphics[width=0.9\columnwidth]{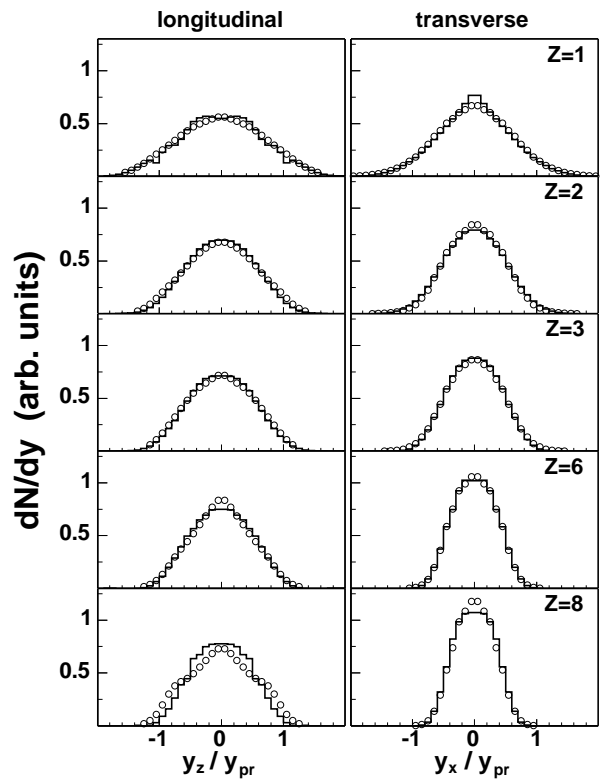}
 \end{center}
  
  \caption{Yield distributions as a function of the scaled longitudinal  (left)
   and transverse (right) rapidity for several fragment species within $Z = 1 -
   8$ for central Au+Au collisions at $150A$ MeV  measured with FOPI (circles)
   and INDRA (histograms). Impact parameter $b_0<0.15$ is selected using $Erat$
   for both cases, the spectra are  normalized to permit an easier comparison
   of their shapes.}

  \label{rapidity}

\end{figure}

Rapidity distributions in longitudinal ($y_z$) and in an arbitrarily fixed 
transverse direction ($y_x$) as obtained with the FOPI and INDRA detectors  for
central Au+Au collisions at $150A$ MeV are shown in fig.~\ref{rapidity}.

\begin{figure}[!htb]
 \leavevmode
 \begin{center}
  \includegraphics[width=0.9\columnwidth]{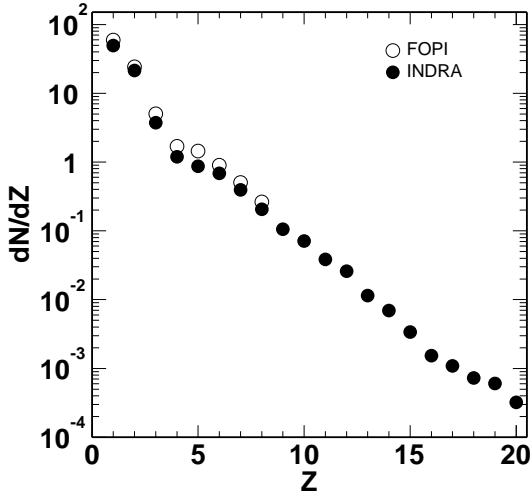}
 \end{center}
  
  \caption{Charge multiplicity distributions in central ($b_0<0.15$) collisions
   of Au+Au at $150A$ MeV. Open circles: FOPI data; closed circles: INDRA
   data.}

  \label{dndz}

\end{figure}

To allow a closer comparison of the shapes  the distributions have been
normalized to the unit area,  individually for each fragment charge. The $Erat$
observable constructed from all detected reaction products except  the particle
of interest was used as impact-parameter selector. In the case of the FOPI
data, the distributions have been reconstructed for the uncovered phase space
and symmetrized with respect to the c.m. rapidity   using two-dimensional
extrapolation methods \cite{reisdorf04a} in the transverse-momentum {\em vs.}
rapidity plane. For $Z=1,2$ these corrections represent less than $10\%$ of the
total yield, for heavier fragments they amount up to $30\%$, leading to
estimated uncertainties of $10\%$ near midrapidity and of $5\%$ for 
$|y|/y_{pr}>0.5$. 
The INDRA distributions have been corrected for the 10\% geometrical 
inefficiency \cite{pouthas95} by multiplying the yields with a factor of 1.11.
The positions of the detected particles and fragments were uniformly randomized
within the active area of the detection modules. For $Z = 1$ the backward c.m.
distribution was used and reflected into the forward hemisphere which is
affected by losses due to  punch-through of energetic particles. For heavier
charges the forward part  was used and symmetrized to profit from the higher
granularity of the detector there and to avoid the higher thresholds affecting
the yields at  backward angles. 

Taking into account the systematic errors (not shown in the figure),  the
agreement of the two independent measurements is very good. This feature is far
from trivial: due to different acceptances, especially for heavier fragments,
the composition of the global event selector cannot be made strictly identical
for the two detectors. Since at this incident energy the difference between
rapidity and velocity is small, one can say that in a naive thermal equilibrium
model, ignoring flow and partial transparency effects, the two kinds of
distributions, longitudinal and transverse, ought to be equal, with the common
variances being a measure of the (kinetic) temperature. Clearly, this is not
the case, the transverse widths are smaller than the longitudinal widths, even
though the selection method, using maximal $Erat$, is definitely biased towards
isolating the event sample (on the 130 mb level) with the largest ratio of
transverse-to-longitudinal variances  (although, as mentioned earlier,
autocorrelations were removed).

The integration over rapidity yields absolute charged-particle  distributions
$dN/dZ$. The results for Au+Au at $150A$ MeV are shown in fig.~\ref{dndz}. For
FOPI, only $Z \leq 8$ yields are available at this energy while the INDRA data
extend over almost 6 orders of  magnitude up to $Z=20$. The observed yields of
heavy fragments are small, however,  only about 2-3\% of the available charge 
is clusterized in fragments with $Z>8$, as expected at this energy where the
c.m. collision energy amounts to four times the nuclear binding energy. With
these 3\% added,  the FOPI data account for 97\% of the total system charge
which is consistent with the  $4\pi$-reconstruction method. The INDRA yields
are systematically lower than FOPI by  between 10\% and 30\%. The lower $Z
= 1$ yield is mainly responsible for the detection of only $80\%$ of the total
system charge with INDRA but similar differences are also observed for larger
$Z$. They are most likely caused by  reaction losses and edge effects in the
detectors which reduce the effective solid-angle coverage if  $Z$
identification is required. The light particle yields may also be affected by
the higher multi-hit probabilities at this incident energy at the upper  end of
the INDRA regime. Extrapolating these observations over the full range of
incident energies studied in this work, one may expect that reaction losses and
the multi-hit probability are considerably reduced at lower incident  energies
for INDRA while the missing yields at large $Z$ in the FOPI case  will be
negligible at higher energies for the mainly central and mid-central 
collisions that are of interest here. 

\begin{figure}[!htb]
 \leavevmode
 \begin{center}
  \includegraphics[width=0.9\columnwidth]{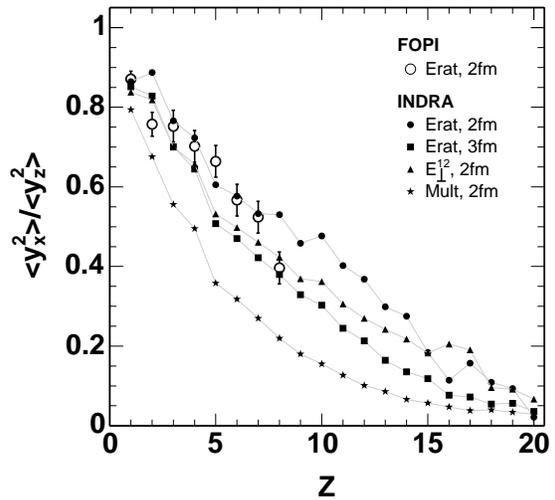}
 \end{center}
  
  \caption{Ratio of transverse to longitudinal variances for central Au+Au
  collisions at $150A$ MeV as measured with the FOPI and INDRA detectors (open
  and filled symbols, respectively). The variances were obtained for scaled
  c.m. rapidities in the  range -1$\leq$y$\leq$1. The chosen selections of
  centrality are indicated  in the legend.} 

\label{vartlz}
\end{figure}

The ratio of the variances of the transverse and longitudinal  rapidity
distributions has recently been proposed as a measure of the degree of stopping
reached in nuclear collisions \cite{reisdorf04}. The ratios obtained for
central Au+Au collisions at $150A$ MeV, after integration over the range of
scaled c.m. rapidity -1$\leq$y$\leq$1, are shown in fig. \ref{vartlz} as a
function of $Z$. The open circles represent the FOPI data with error bars which
include the systematic uncertainty of the reconstruction procedure. The INDRA
data are shown for $Z \leq 20$ and for four different impact  parameter
selections as indicated in the figure. 

The largest ratios 0.8 to 0.9 are observed for light charged particles  ($Z
\leq 2$). With increasing fragment $Z$, the ratios decrease continuously  to
values of $< 0.1$ near $Z = 20$. In the common range of fragment $Z$  and for
the same impact-parameter selection ($Erat, b \leq 2$~fm), the ratios measured
with FOPI and INDRA are in good agreement. The selection with $Erat$ and $b
\leq 3$~fm yields slightly smaller ratios as expected which, however, are
similar to those obtained with $E_{\perp}^{12}$.  Large transverse momenta of
light charged particles and of fragments are apparently correlated.
Autocorrelations are not present here because the particle of interest is
removed from the impact parameter selector (see previous section). The smallest
ratios of variances are obtained for selections according to multiplicity.

\begin{figure}[!htb]
 \leavevmode
 \begin{center}
  \includegraphics[width=0.9\columnwidth]{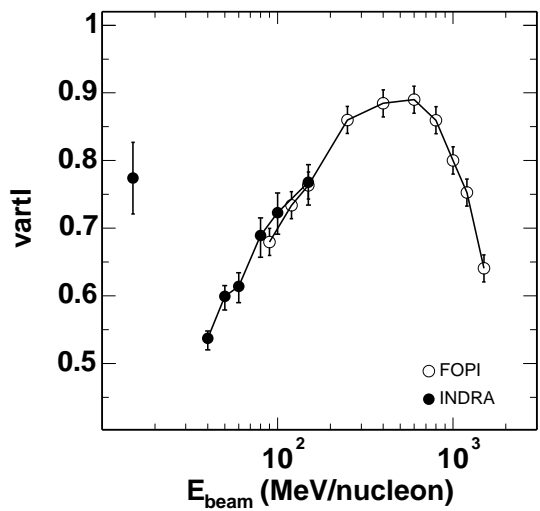}
 \end{center}
  
  \caption{Excitation function of the degree of stopping, $vartl$, in central
  Au+Au collisions ($b\leq2$~fm) obtained from the FOPI (open circles) and
  INDRA (dots) measurements. The result at $15A$ MeV corresponds to a less
  central selection ($b\leq5$~fm).}

\label{vartl}
\end{figure}

The trends as observed as a function of $Z$ suggest that the heavier fragments,
even in rather central collisions, experience less stopping than lighter ones
and keep a strong memory of the entrance channel motion.  Their transverse
momenta seem to be, nevertheless, generated in collisions  involving nucleons
or light clusters as evident from the correlation with  $E_{\perp}^{12}$. The
momenta of struck nucleons absorbed in a cluster or the recoil momenta of
nucleons knocked out from a cluster both contribute to their final momenta.
Their relative weight will be smaller in larger fragments, consistent with the
observed $Z$ dependence.  Overall, these observations are clearly in
contradiction to the assumption of global equilibrium including the kinetic
degrees of freedom. Qualitatively, they agree with the predictions of quantum
molecular dynamics calculations for fragment production in this energy range
\cite{zbiri}.

A global observable to describe stopping, $vartl$, has been introduced  in ref.
\cite{reisdorf04}. It is defined as the ratio of the transverse over
longitudinal variances of the summed and $Z$-weighted rapidity distributions. 
The excitation function of  this observable is presented in fig.~\ref{vartl}
for central Au+Au collisions ($b\leq2$~fm) and for the full energy range
covered in the FOPI and INDRA experiments.

The FOPI results have been obtained with the $Erat$ selection. The data have
recently been reanalyzed by taking additional small corrections due  to energy
losses in structural parts of the detector into account. At the lowest three
energies, this has led to an increase of $vartl$ by up to about 10\% compared
to the data  published in \cite{reisdorf04}, while for the other energies the
results are unaffected. The measured range of fragments  extends up to $Z =
6,8,8$ for $E=90A$, $120A$, $150A$ MeV, respectively. The contribution of
heavier fragments to the $vartl$ observable has been estimated by extrapolating
their weights and variance ratios to higher $Z$. At $90A$ MeV this correction
amounts to about 8\%. The errors given in the figure are systematic and mainly 
reflect the uncertainty of the reconstruction procedure.

For the INDRA central event samples, the light-particle transverse energy
E$_{\perp}^{12}$ has been used to select $b\leq2$~fm for $40A$ to $150A$ MeV
and $b\leq5$~fm for the data sample at $15A$ MeV with low statistics,
originally only taken for calibration purposes. For the charge-weighted average,
fragments up to $Z=60$ have been included. Except for the result at $15A$ MeV,
the error bars correspond to the variation of $vartl$ for centrality selections
within $b\leq1.5$~fm (upper end of the error bar) and $b\leq2.5$~fm (lower end)
and thus represent the systematic  uncertainty associated with the impact
parameter determination. The statistical errors are below the percent level,
except at $15A$ MeV, where they are the main contribution to the error shown in
the figure.

The obtained excitation function of stopping is characterized by a broad
plateau extending from about $200A$ to $800A$ MeV with fairly rapid drops above
and below. The highest value reached by $vartl$ is about 0.9. With the INDRA
data, the reduction of stopping at lower incident energies is followed down to
$40A$ MeV. In the overlap region, a very satisfactory agreement within errors
is observed.  The measurement at $15A$ MeV suggests that stopping goes  through
a minimum at or below $40A$ MeV.

It is clear that only a dynamical theory will be able to reproduce this
excitation function. Using the relativistic Boltzmann-Uehling-Uhlenbeck (RBUU)
transport model, an analysis of the combined FOPI stopping and flow (see later) data was  recently presented \cite{gaitanos05}. The input to computer codes
implementing transport theoretical models are the nuclear mean field $U$ (or
EoS) and the nucleon-nucleon ($nn$) cross sections $\sigma_{nn}$. Although both
are not independent in a consistent theory, it is useful to consider their
effects separately. In general one finds that the cross sectional part is
dominant over the mean field part for a quantitative account of the observed
incomplete stopping: note that if global equilibrium, or even local equilibrium
(ideal hydrodynamics), were valid cross sections would be irrelevant. Starting
at the low energy end one qualitatively expects, when raising the energy, that
the increasingly repulsive mean field (due to increasing compression) and the
drop in Pauli-blocking of final and intermediate states in $nn$ scattering (due
to the increasing initial rapidity gap) conspire to raise rapidly the
generation of transverse energy at the expense of the longitudinal energy. At
the higher energy end (say beyond $1A$ GeV) again both aspects (mean field
and collisions) more or less may add up to make the drop faster. At $1.5A$
GeV roughly one quarter of the nucleons are excited to a resonant state. The
opening up of nucleonic degrees of freedom may lead to a softening of the EoS.
On the other hand the in-medium Dirac masses $M^*_D$ are predicted to drop
substantially in covariant theories~\cite{brockmann90, vandalen05, mishra04}, a
fact that will seriously modify the phase-space and kinematical factors
influencing the elementary cross sections~\cite{malfliet88, fuchs01,
larionov03}. The calculations of ref.~\cite{gaitanos05} suggest that these
in-medium modifications of $\sigma_{nn}$ are indeed necessary to reproduce the
observed stopping.

Besides the `global' information shown in fig.~\ref{vartl} the `particle
differential' information reveals additional information on the stopping
mechanisms. Figure~\ref{vartlz} shows that the partial transparency is
predominantly experienced by the heavier fragments, which presumably have
survived because their constituent nucleons have suffered a less violent
average collision history. This feature is also observed at the high energy
end, although the `heavy fragment' role is played there by mass $A=2-4$
ejectiles~\cite{fopi-tobe}. Restricting the stopping observable to the lightest
species at the various incident energies, one obtains higher $vartl$ values and
flatter excitation functions. The combined role of the mean field and of
in-medium modified cross sections will be picked up again in section
\ref{sec:correlation} where the flow information will be added to the analysis.

\section{Flow, reaction plane and corrections}
\label{sec:corrections}

Originally, the directed flow has been quantified by measuring the in-plane
component of the transverse momentum \cite{danielewicz85} and the elliptic flow
by parametrizing the azimuthal asymmetries using the Fourier expansion fits
\cite{gutbrod89, demoulins90}. More recently, it has been proposed
\cite{voloshin96} to express both, directed and elliptic flow in terms of the
Fourier coefficients ($v_{1}$ and $v_{2}$, respectively) and also to
investigate the higher flow components. The coefficients $v_{n}$ are obtained
by means of the Fourier decomposition \cite{voloshin96, ollitrault97,
poskanzer98} of the azimuthal distributions measured with respect to the true
reaction plane:  
\begin{equation} 
% \frac{dN}{d(\phi-\phi_R)} \propto 1+2 \sum_{n\geq1} v_n \cos n(\phi-\phi_R) 
\frac{dN}{d(\phi-\phi_R)} = \frac{N_0}{2\pi} 
\left( 1+2 \sum_{n\geq1} v_n \cos n(\phi-\phi_R)\right)
\label{eq:defvn} 
\end{equation}

\noindent with $\phi_R$ being the azimuth of the latter. In general, the
coefficients $v_{n} \equiv \langle\cos n(\phi-\phi_R)\rangle$ may depend on the
particle type, rapidity $y$ and the transverse momentum $p_{\rm T}$. 

The standard methods of measuring flow can be split into those using explicitly
the concept of the reaction plane \cite{danielewicz85, voloshin96,
ollitrault97, poskanzer98} and those based on the two-particle azimuthal
correlations \cite{wang91}.  Still other methods have been proposed recently,
satisfying the needs of high energy experiments: the `cumulant' methods
\cite{borghini01a, borghini01, borghini02a} using multi-particle correlations
and the method based on the Lee--Yang theory of phase transitions
\cite{bhalerao03, bhalerao04}. The latter is expected to perform well above
about $100A$ MeV \cite{bhalerao03}, while the three-particle variant of the
`cumulant' method is claimed to be useful for extracting  $v_{1}$ coefficients
at energies near the balance energy and in the ultrarelativistic regime
\cite{borghini02a}.  However, because the correlation methods require  high
event multiplicities and high-statistics data, and because the correlation
between a particle and the flow vector is usually much stronger than that
between two particles \cite{borghini02}, the reaction plane methods are still
more commonly used at intermediate energies. They have also been applied in the
present case.

Since detectors do not allow to measure the angular momenta and spins of the
reaction products, the orientation of the reaction plane can only be estimated
using the momenta. The resulting azimuthal angle, $\phi_E$, has a finite 
precision, and the measured coefficients $v_{n}^{meas}$ are thus biased. They
are related to the true ones through the following expression
\cite{ollitrault97}:
\begin{equation}
v_n^{meas} \equiv \langle\cos n(\phi-\phi_E)\rangle = 
v_n \langle\cos n\Delta\phi\rangle
\label{eq:defcorr}
\end{equation}

\noindent where the average cosine of the azimuthal angle between the true and
the estimated planes, $\langle\cos n\Delta\phi\rangle \equiv \langle\cos
n(\phi_R-\phi_E)\rangle$, is the required correction (also referred to as
`event plane resolution' or just `resolution') for a given harmonic. Note that,
since the true values of flow are obtained by dividing by the average cosine,
they are always larger than the measured ones.

The literature offers many different methods to estimate the reaction plane,
like the flow-tensor method \cite{gyulassy82}, the fission-fragment plane
\cite{tsang84a}, the flow Q-vector method \cite{danielewicz85}, the transverse
momentum tensor \cite{ollitrault93} (also called `azimuthal correlation'
\cite{wilson92}) method or others \cite{fai87}. 

Among them, the Q-vector method has received special attention. Originally,
the Q-vector has been defined as a weighted sum of the transverse momenta of
the measured $N$ reaction products \cite{danielewicz85}:
\begin{equation}
% \vec{Q} = \sum_{\stackrel{\scriptstyle i=1}{i\neq POI}}^{N}\omega_i \vec{p}_i^{\perp}
% \vec{Q} = \sum_{i\neq POI}^{N}\omega_i \vec{p}_i^{\perp}
\vec{Q} = \sum_{i=1}^{N}\omega_i \vec{p}_i^{\perp}
\label{eq:defqvect}
\end{equation}

\noindent with the weights $\omega$ chosen to be +(-)1 for reaction products in
the forward (backward) c.m. hemisphere and with the possibility to  exclude
the midrapidity zone. The choice of the optimal weights is discussed in
\cite{ogilvie89a, tsang91, danielewicz95, poskanzer98, borghini01}. Definition
(\ref{eq:defqvect}) can be extended to Q-vectors built from higher harmonics
\cite{poskanzer98}, thus {\em e.g.} allowing to profit from strong elliptic
flow, when applicable. Usually in the flow studies, the POI is excluded from
the sum in (\ref{eq:defqvect}) to avoid autocorrelations. This does not concern
the corrections, since the sub-events (see below) do not share particles.

The corrections for the reaction-plane dispersion can be obtained using various
methods \cite{danielewicz85, danielewicz88, ollitrault93, ollitrault95,
voloshin96, ollitrault97, ollitrault98, poskanzer98, voloshin99, borghini02}.
What they all have in common, is the underlying assumption of  the
applicability of the central-limit theorem. In most of these methods the
correction is searched for using the sub-event method \cite{danielewicz85},
which consists in splitting randomly each event into two equal-multiplicity
sub-events and getting the correction from the distribution of the relative
azimuthal angle, $\Delta\Phi_{12}$, between their individual Q-vectors
(`sub-Q-vectors'). This is done either by using the small angle expansion
\cite{danielewicz85} or by fitting with a theoretical distribution
\cite{ollitrault97}. Instead of fitting the angular distributions one can
alternatively fit the distributions of the magnitude of the total Q-vector
itself \cite{voloshin96, poskanzer98}.

Assuming the Gaussian limit, ref. \cite{ollitrault97} gives an analytical
formula for the distribution of $\Delta\Phi_{12}$ for the case that the
distributions of sub-Q-vectors are independent and isotropic around their mean
values. In refs. \cite{voloshin96, poskanzer98} one can find the formulae
relevant for the distributions of the magnitude of the Q-vector.

These methods proved their usefulness for correcting measured flow values at
higher energies (see {\em e.g.} \cite{danielewicz85, barrette94, aggarwal97,
andronic01, andronic05}) which fulfill the high multiplicity requirement. They
are, however, not adequate for the intermediate energy reactions, below about 
$100A$ MeV, where the particle multiplicities are lower and the events are
characterized by a broad range of masses of the reaction products. Here, the
applicability of the central limit theorem for devising the corrections is less
obvious.

\begin{figure}[!htb]
 \leavevmode
 \begin{center}
  \includegraphics[width=0.95\columnwidth]{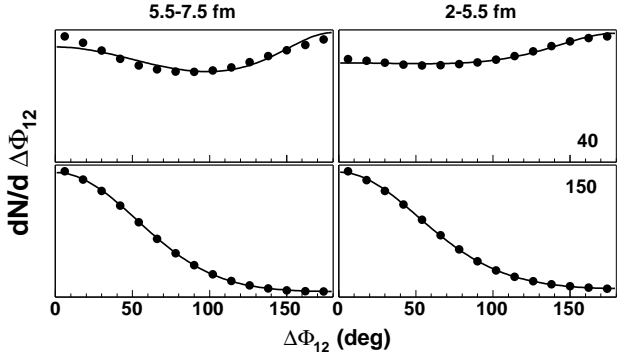}
 \end{center}
  
  \caption{Distributions of relative angle between reaction planes for two 
  random sub-events (INDRA data, dots)
  and fits (lines) using the integrated product of bivariate normal 
  distributions (Eq. \protect\ref{eq:q1q2distr}) for centralities
  $b \simeq 5.5-7.5$~fm (left) and $b \simeq 2-5.5$~fm (right) and incident energies
  $40A$ MeV (top panels) and $150A$ MeV (bottom panels).}

\label{fig:ang_dis}
\end{figure}

Figure~\ref{fig:ang_dis} illustrates the difficulties one encounters at
intermediate energies. It shows the experimental distributions of 
$\Delta\Phi_{12}$ as measured with  the INDRA detector for the Au+Au reaction
at $40A$ (top) and $150A$  (bottom) MeV and for two intermediate centrality
bins. The lines represent the fits obtained with the method described briefly
below. The standard method \cite{ollitrault97} can, in principle, be used to
derive the corrections for energies down to about $80A$ MeV, however it fails
to describe distributions like  those at $40A$ MeV with double maxima or maxima
at backward angles, which reflect the presence and importance of the in-plane
enhancement and of the correlation between sub-events.

Since at low and intermediate energies the sub-events are expected to be
strongly correlated \cite{ollitrault95} and the distributions of the Q-vector
no longer necessarily isotropic \cite{ollitrault97}, we have extended the
method of Ollitrault \cite{ollitrault97} by explicitly taking into account
these two effects in the theoretical distribution of the sub-Q-vectors. The new
method relies on the assumption of the Gaussian distribution of the flow
sub-Q-vectors. This assumption has been verified to hold even at $40A$ MeV,
except for very peripheral collisions, by performing tests with the
CHIMERA-QMD  model in which angular momentum is strictly conserved
\cite{method}. 

The form of the joint probability distribution of the random sub-Q-vectors has
been searched for following the method outlined in appendix A of
\cite{borghini02}, by imposing the constraint of momentum conservation on the
$N$-particle transverse momentum distribution and using the saddle-point
approximation.

The resulting distribution has the form of a product of two bivariate Gaussians:
\[
\frac{d^{4}N}{d\vec{Q}_1d\vec{Q}_2} = 
\frac{1}{\pi^2\sigma_{sx}^2 \sigma_{sy}^2 (1-\rho^2)} 
\cdot \exp \left[ \vphantom{\frac{(Q_{1x})^2}{(Q_{1x})^2}} \right.
\]
\[
-\frac{(Q_{1x}-\bar{Q}_s)^2+(Q_{2x}-\bar{Q}_s)^2 -2\rho\;
(Q_{1x}-\bar{Q}_s)(Q_{2x}-\bar{Q}_s)}{\sigma_{sx}^2 (1-\rho^2)}
\]
\begin{equation} 
-\frac{Q_{1y}^2+Q_{2y}^2-2\rho\; Q_{1y}Q_{2y}}{\sigma_{sy}^2 (1-\rho^2)} 
 \left.\vphantom{\frac{(Q_{1x})^2}{(Q_{1x})^2}}\right] 
\label{eq:q1q2distr} 
\end{equation}

% \[
% \frac{d^{4}N}{d\vec{Q}_1d\vec{Q}_2} = 
% \frac{1}{\pi^2\sigma_{sx}^2 \sigma_{sy}^2 (1-\rho^2)} 
% \]
% \begin{equation} 
% e^{-\frac{(Q_{1x}-\bar{Q}_s)^2+(Q_{2x}-\bar{Q}_s)^2 -2\rho
% (Q_{1x}-\bar{Q}_s)(Q_{2x}-\bar{Q}_s)}{\sigma_{sx}^2 (1-\rho^2)}}
% \label{eq:q1q2distr} 
% \end{equation}
% \[
% e^{-\frac{Q_{1y}^2+Q_{2y}^2-2\rho Q_{1y}Q_{2y}}{\sigma_{sy}^2 (1-\rho^2)}}
% \]

\noindent where we followed the convention of \cite{ollitrault97} of including
the $\sqrt 2$ in $\sigma$; the subscripts {\em 1, 2} refer individually  and
{\em s} generally to sub-events;  the subscripts {\em x} and {\em y} refer to
the in- and out-of-plane direction, respectively. This distribution differs
from those proposed in  \cite{ollitrault97, ollitrault95, borghini02} in that
it combines all three effects that influence the reaction plane dispersion at
intermediate energies, namely the directed flow (through the mean in-plane
component $\bar{Q}_s$ or the resolution parameter $\chi_{s} \equiv
\bar{Q}_s/\sigma_{sx}$ \cite{ollitrault97}), the elliptic flow (through the
ratio $\alpha \equiv \sigma_{sx}/\sigma_{sy}$) and the correlation between the
sub-events \cite{ollitrault95} (through the correlation coefficient $\rho
\in[-1,1]$). It reduces to the one of \cite{ollitrault97} for $\alpha = 1$ and
$\rho = 0$, and to the one of \cite{borghini02} for $\alpha = 1$. In deriving
eq. (\ref{eq:q1q2distr}) it was assumed that the in- and out-of-plane
correlation coefficients are equal.

Making the division into sub-events random ensures that the distributions of
the sub-Q-vectors are equivalent, in particular they have the same mean values
and variances. Since the total-Q-vector is the sum of the sub-Q-vectors,
$\vec{Q}=\vec{Q}_1+\vec{Q}_2$, one finds the following relation between the
resolution parameter obtained from the distribution of the Q-vector, $\chi$,
and that obtained from the distribution of sub-Q-vectors, $\chi_{s}$:
\begin{equation}
\chi = \chi_{s} \sqrt{2/(1+\rho)}
\label{eq:chi}
\end{equation}

\noindent Relation (\ref{eq:chi}) shows how the correlation between sub-events
influences the reaction plane resolution. In particular, it indicates that the
resolution improves in case the sub-events are anti-correlated ($\rho<0$),
which is predicted to be the case below about $150A$ MeV except for very
peripheral collisions \cite{method}. 

As in \cite{ollitrault97}, the joint probability distribution 
(\ref{eq:q1q2distr}) is used after integrating it over the magnitudes  of the
sub-Q-vectors and one angle, leaving the $\Delta\Phi_{12}$ as the only
independent variable. Unlike in \cite{ollitrault97}, the resulting distribution
can not be presented in an analytical form. It depends on 3 parameters
($\chi_{s}, \alpha, \rho$) which can be obtained from fits to the experimental
or model data. The quality of the obtained fits is very good, even in the
non-standard cases encountered at low energies (fig.~\ref{fig:ang_dis}).

\begin{figure}[!htb]
 \leavevmode
 \begin{center}
  \includegraphics[width=0.95\linewidth]{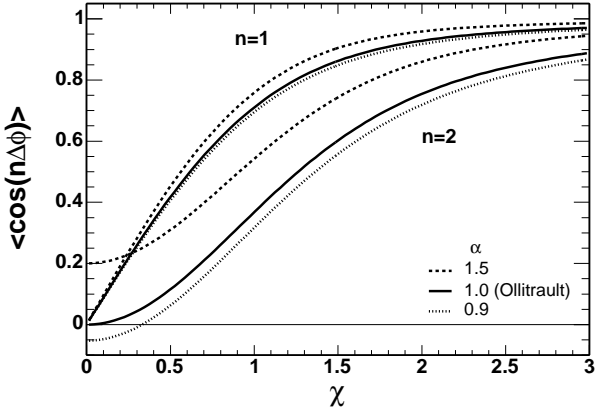}
 \end{center}
  
  \caption{Corrections for the first (3 upper curves) and the second (3 lower
  curves) harmonic as a function of the resolution parameter, $\chi$, for
  different aspect ratios, $0.9\leq\alpha\leq1.5$, covering approximately the
  range of its variation.} 

  \label{fig:olli15}

\end{figure}

The corrections for the n-th harmonic $v_{n}$, depending now on $\chi$ and
$\alpha$, can be calculated (also numerically) as the mean values of the $\cos
n\Delta\phi$ obtained over the total-Q-vector distribution, in a similar way as
in \cite{ollitrault97}.  
Figure \ref{fig:olli15} shows how the elongation of the Gaussian ($\alpha$), 
resulting from elliptic flow,  modifies the corrections for the first two
harmonics. It demonstrates that the in-plane emissions ($\alpha>1$) enhance
slightly the resolution for $v_1$ and considerably for $v_2$ -- even in the
absence of the directed flow. On the other hand, squeeze-out ($\alpha<1$)
deteriorates the resolution. The figure, in particular, shows that the
correction can change the sign of elliptic flow in the case of small directed
flow and squeeze-out. 

Since the correlation between the sub-events increases at the lower energies,
the knowledge of the correlation coefficient $\rho$  becomes crucial. Estimated
values, obtained from model calculations, can be useful as constraints for
$\rho$ in the fitting procedure. For example, the CHIMERA-QMD calculations
predict $\rho$ to be around -0.43 for $40A$ MeV and $2<b<8$~fm and about -0.2
at $150A$ MeV. Alternatively, mean values of some rotational invariants which
are derived from the measured data can be used to reduce the number of fit
parameters and to constrain the  fitting routine to  search for a conditional
minimum \cite{method_iwm}.

Instead of fitting the azimuthal distributions one can express the probability
distribution (\ref{eq:q1q2distr}) in terms of the components of one of the
sub-Q-vectors in the reference frame of the other, or in terms of the absolute
values of the sum and of the difference of the sub-Q-vectors. The corresponding
2-dimensional experimental distributions can then be fit using such formulae.
The method of fitting the distributions of components of the sub-Q-vector has
been found sensitive enough to perform well without additional constraints.

\begin{figure}[!htb]
 \leavevmode
 \begin{center}
  \includegraphics[width=0.9\columnwidth]{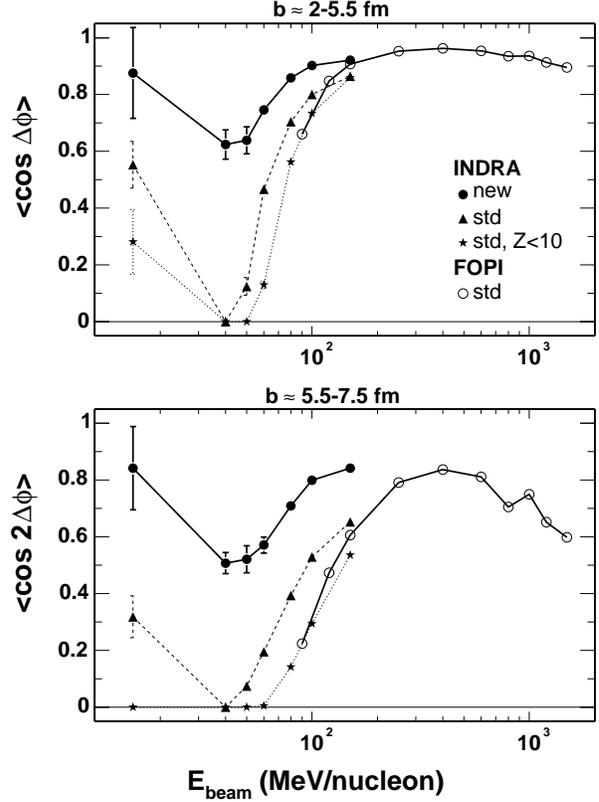}
 \end{center}
  
  \caption{Corrections for the first  harmonic and impact parameter $b
  =2-5.5$~fm (top) and for the second  harmonic and $b =5.5-7.5$~fm (bottom)
  used in the flow analysis of the  Au+Au reaction. The corrections for the
  INDRA data (full symbols)  are shown  as obtained with the new method (dots,
  see text), with the standard (std) method (ref.~\protect\cite{ollitrault97},
  triangles), and with the standard  method and the restriction $Z<10$ (stars).
  Open circles represent the results for the FOPI data obtained with the
  standard method.}

  \label{fig:corrections}

\end{figure}

The corrections obtained using various methods are presented in 
fig.~\ref{fig:corrections}. They are close to one, independent of the method,
for the range of higher incident energies ($E > 100A$ MeV) where the directed
flow is large and the reaction plane well defined by the high-multiplicity
distribution of detected particles. At around $50A$ MeV, they go  through a
minimum and depend strongly on the chosen method.  The FOPI flow results, as
published in refs.~\cite{andronic01,andronic05},  have been corrected using the
standard method, excluding the midrapidity  region of $\pm0.3$ of the scaled
c.m. rapidity from the Q-vector to improve the resolution. The corrections used
here for the INDRA data are obtained with the new method in two ways, by 
fitting the azimuthal distributions and by fitting the distributions of
components of the sub-Q-vectors. The mean values  are given in
fig.~\ref{fig:corrections} (full circles) with error bars representing the
systematic uncertainty as given by the difference of these results. At $15A$
MeV, the statistical errors dominate. Even at their minima,  the corrections
are not smaller than 0.6 and 0.5 for directed and  elliptic flow, respectively,
indicating that the measured flow values will increase, after applying the
corrections, by no more than about a factor of two. 

For a comparison of the different methods and of their applicability, also the
corrections according to the standard method of Ollitrault \cite{ollitrault97}
have been determined. This corresponds to fixing the parameters $\alpha=1$ and
$\rho=0$ in the new method. Near $50A$ MeV, the results are close to zero which
would require nearly infinitely large corrections (triangles in
fig.~\ref{fig:corrections}). The figure, furthermore, shows the same
corrections according to the standard method as obtained for the FOPI case
(circles). They are very similar and, in the overlap region, virtually
identical to the result for INDRA if the limit $Z<10$ of the FOPI acceptance is
applied in the INDRA case (stars). This very close agreement is not unexpected 
because good agreement was already observed for the uncorrected flow data
obtained with the two detection systems \cite{lukasik04, lukasik05,
andronic05}. The standard method, nevertheless, fails below about $80A$ MeV. As
mentioned above, the independent, isotropic Gaussian  approximation is no
longer confirmed by satisfactory fits of the experimental distributions.

Several additional observations and comments can be made. Comparing the results
of the new and standard methods (filled circles and triangles) shows a dramatic
improvement of the resolution obtained by taking the effects of the correlation
between the sub-events and of the in-plane enhancement into account. It should
be stressed, that both these effects are responsible for the finite correction
for the directed flow at $40A$ MeV, near the expected balance energy
\cite{magestro00}. Non-vanishing resolution, suggested by the new method,
indicates that even here the reaction plane can be defined, and apparently, 
questions the occurrence of the 'global' balance, which otherwise would
manifest itself with the vanishing of $\langle\cos \Delta\phi\rangle$. However,
the finiteness of the corrections around $40A$ MeV may also partially result
from the incompleteness of the experimental data and from the mixing of events
with different centralities, which may add up to mask the signal of the
'global' balance. The fitting procedure yields relatively accurate results for
the corrections for the first two harmonics in case of the complete results of
the simulations ({\em e.g.} 2-5\% accuracy for $40A$ MeV and 0.2-0.4\% for
$150A$ MeV and $4<b<8$ fm \cite{method}), and in particular, is able to reveal
the signal of the 'global' balance, but in the case of experimental data it
will certainly return some effective corrections biased by the experimental
uncertainties and inefficiencies. The effects of the latter may not necessarily
drop out by applying eq. (\ref{eq:defcorr}) to correct the measured
observables, but may require additional corrections. At the higher energies,
the results of the standard and the new methods approach each other but, in the
overlap region of the FOPI and INDRA experiments, the differences are still
significant, and need further investigation. 

Comparing the less and the more complete data sets (stars and triangles,
respectively) shows that the resolution improves with the completeness of the
data. Triangles represent INDRA events with at least 50\% of the system charge
collected to which an additional single fragment carrying the missing momentum
and charge was added. This artificial completion of events was found important
for peripheral collisions where, due to the energy thresholds, the heavy
target-like fragment is always lost. The distributions of the relative angle
between sub-events become then narrowly peaked at small relative angles which
improves the resolution of the reaction plane.

However, it is not only the reaction plane correction that relies on the
completeness of the measured data. Also the measured $v_{n}^{meas}$ parameters
are affected by the non-isotropic loss of particles due to multi-hits
(INDRA) or unresolved tracks in high-track-density regions (FOPI). A rough
estimate of the correction \cite{method} due to multi-hit losses for $v_{2}$ can
be obtained, for segmented detectors like INDRA, by using the unfolded 'true'
in- and out-of-plane multiplicities and calculating the true and measured mean
$v_{2}$ by integrating the azimuthal distribution (\ref{eq:defvn}) over the in-
and out-of-plane quadrants. The 'true' multiplicities can be estimated using
the calculated ({\em e.g.} in a way similar to that of ref.
\cite{vanderwerf78}) or simulated (using the detector filter and the model
data) multiplicity response function specific for a given detector. An 
analogous procedure can be applied also for $v_{1}$; however, due to the lack 
of the forward-backward center of mass symmetry of the detector, 
the results may be less accurate. 
The flow parameters obtained from the INDRA data presented in the 
next section have been additionally corrected for the multi-hit losses using
the above procedure. For $v_{1}$, these additional 
corrections vary from about
7\% at $40A$ MeV to about 33\% at higher energies for $Z=1$ 
and do not exceed
15\% for $Z=2$. For $v_{2}$ and $Z=1$ they increase 
from about 18\% at $40A$
MeV to about 36\% at $100A$ MeV and about 70\% at $150A$ MeV, for the centrality
bins in question. Within this simple procedure, the corrections depend
essentially on the average of the in- and out-of-plane multiplicities and only
weakly on their difference, that is why the corrections basically increase with
the increasing multiplicity (thus with the centrality and incident energy).
This explains the large correction factor at $150A$ MeV. Nevertheless, since
$v_{2}$ is small at this energy, the absolute change of the measured value due
to the correction is small compared to that at lower energies.

%float Mcor_z1[7]  = { 1.070163, 1.252491, 1.282414, 1.306540, 1.325279, 1.329954, 1.302434};
%float Mcor_z2[7]  = { 1.097852, 1.132471, 1.130572, 1.126057, 1.123049, 1.116793, 1.097628};

%float  Mcorf[7]   = { 1.184612, 1.295328, 1.319249, 1.322444, 1.343530, 1.363769, 1.692583};

% Generally, one may remark that correcting for the experimental dispersion
% of the reaction plane becomes a delicate task at energies below about 
% 100 $A$MeV. The corrections may change the measured flow values by a large
% factor, mainly also because of the 
% smallness of directed flow near 50~$A$MeV. The uncertainty for,
% e.g., $v_2$ which is not small there becomes correspondingly
% large. The corrected 'true' flow values are of interest from the 
% theoretical point by permitting the direct comparison with the model 
% predictions. In problematic cases, this may be tested by using a realistic
% filter routine and performing the comparison on the level of uncorrected 
% observables.

Generally, one may remark that, at energies below about $100A$ MeV, 
devising the
corrections becomes a delicate task. The corrections are no longer those in the
usual sense, say, of a few percent. Depending on the method, they may change
the measured results by a large factor, mainly because of the  smallness of
directed flow around $40A-50A$ MeV. The accuracy relies in addition on the
completeness of the data. Flow data free of reaction plane dispersions are,
nevertheless, very desirable 
%The reaction plane dispersion free values of flow
%are, however, of great importance 
since they allow to compare the results
obtained with different detectors. They are also of great interest from the
theoretical point of view, by permitting the direct comparison with the model
predictions. In problematic cases, however, detailed filtering of the model
results and treating them with the experimental type of analysis may still be
%desirable, 
necessary, if not for the direct comparison on the level of uncorrected 
observables, then for the reliable estimate of the systematic uncertainties
associated with the correction scheme.

The effects of momentum conservation, distortions due to removal of the
particle of interest (expected to be important at low multiplicities
(energies)) and possible corrections to the reaction plane resolution due
to the detector inefficiencies (missing part of the Q-vector) remain a subject
for future study. 

\section{Directed and elliptic flow}
\label{sec:flow}

\begin{figure}[!htb]
 \leavevmode
 \begin{center}
  \includegraphics[width=0.9\columnwidth]{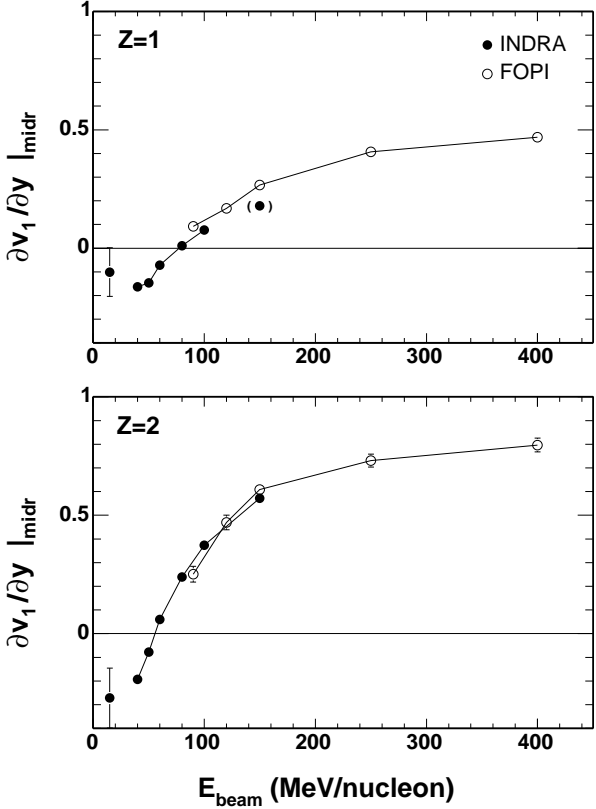}
 \end{center}
  
  \caption{Slopes of directed flow $\partial v_{1}/\partial y$ for $Z=1$ (top)
  and $Z=2$ (bottom) particles integrated  over $p_{\rm T}$ for mid-central
  collisions (2--5.5 fm). The open and filled symbols represent the FOPI
  \protect\cite{andronic01} and the INDRA data, respectively. The uncertainty
  at $15A$ MeV is mainly statistical. The INDRA point, in brackets, at $150A$
  MeV in the top panel is biased due to  experimental inefficiencies for $Z=1$
  at this energy.}

\label{fig:v1corr}
\end{figure}

The rapidity dependence of the slope of the directed-flow $\partial
v_{1}/\partial y$ at midrapidity for $Z=1$ and 2 particles, integrated over
transverse momentum, is shown in fig. \ref{fig:v1corr}. The INDRA data is
combined with the FOPI data (published for $Z=2$ in \cite{andronic01}), both
measured for mid-central collisions with impact parameters of 2--5.5 fm and
shown after correcting for the  reaction plane dispersion. The FOPI data has
been corrected using the method of \cite{ollitrault97} while the INDRA data has
been corrected using the method outlined in sect. \ref{sec:corrections}. In
both data sets the reaction plane has been reconstructed using the Q-vector
method with the weights $\omega=sign(y_{cm})$, excluding the POI. In case of
the FOPI data the midrapidity region of $\pm0.3$ of the scaled rapidity has
been excluded from the Q-vector to improve the resolution. The INDRA data has
been corrected for the effects of momentum conservation \cite{ogilvie89}. In
both cases linear fits have been performed in the range of $\pm 0.4$ of the
scaled c.m. rapidity, except for the $15A$ MeV data where the range of $\pm
0.55$ was used.

The excitation function of the slope of the directed flow at midrapidity for
$Z=1$ changes sign around $80A$ MeV. The apparent minimum around $40A$ MeV is
mostly suggested by the $15A$ MeV data point and should be confirmed by other
measurements. The FOPI data has been additionally corrected for the effects of
unresolved tracks in the in-plane high track density region. This correction
influences also the slope of the $v_1$ rapidity distribution, increasing it by 
up to 15\% for $Z=1$ and up to 5\% for $Z=2$. The INDRA results have been
corrected for the effect of multi-hit losses (see sect. \ref{sec:corrections}).
The still apparent discrepancy between the INDRA and FOPI results at $150A$ MeV
can be partially attributed to the losses of $Z=1$ particles due to
punch-through effects in the INDRA detector at high energies.  Up to 10\% of
the difference may also come from the different methods used for correcting 
the reaction plane dispersions (see  fig.~\ref{fig:corrections}).

For $Z=2$, the slope of $v_1$ is seen to rise monotonically with energy over
the full range of 15 to 400 MeV per nucleon which is  covered by the two
experiments. Here, the agreement in the overlap  region is slightly better
reflecting the better efficiency of the INDRA detector for $Z=2$ particles. 
The trends observed for the uncorrected data \cite{lukasik05} for $v_1$ are
preserved. Unlike in ref. \cite{magestro00}, the excitation function does not
show a clean signature of a minimum (see ref. \cite{lukasik05} for discussion).
It changes sign between 50 and 60 MeV per nucleon, in agreement with the
extrapolated values of the balance energy, $E_{bal}$, obtained from the
higher energy measurements \cite{zhang90, partlan95, crochet97}. 

Negative flow is observed not only for $Z=1,2$ (fig. \ref{fig:v1corr}) but with
even larger slopes also for other light fragments.
This intriguing phenomenon has already been reported for the
lighter systems  $^{40}$Ar + $^{58}$Ni, $^{58}$Ni + $^{58}$Ni, and $^{129}$Xe +
$^{\rm nat}$Sn,  provided the `1-plane-per-particle' method was used for
estimation of the reaction plane \cite{cussol02}.  For these systems, a balance
energy has been  determined by associating it with the minima of the
approximately parabolic  excitation functions of the flow parameter which, in
the cases of  $^{40}$Ar + $^{58}$Ni and $^{58}$Ni + $^{58}$Ni, appeared at
negative flow values. Negative flow values of light reaction products can
indeed be measured experimentally, provided the detector is able to measure
`quasi-complete' events, including the heavy fragments. Then, the observed
anti-flow of light products is measurable relative to the reaction plane fixed
and oriented by the heavy remnants.

A possible scenario of the anti-flow has been proposed for the lighter systems
in \cite{cussol02}, and for the heavy systems, emphasizing the role of the
strong Coulomb field, in \cite{lukasik05}. Despite the appeal of a globally
defined balance energy, it is worth noticing that directed flow apparently
never vanishes completely. It was shown with BUU calculations that at the
balance energy the flow cancellation results from a complex transverse
momentum  dependence and that the flow pattern is influenced by EoS and 
$\sigma_{nn}$ \cite{li99}. The presently available differential data, measured
by FOPI down to  90 AMeV \cite{andronic01} suggest that the change of sign of
$v_1$ is dependent, in addition to transverse momentum, also on particle type
and  rapidity. 

\begin{figure}[!htb]
 \leavevmode
 \begin{center}
  \includegraphics[width=0.9\columnwidth]{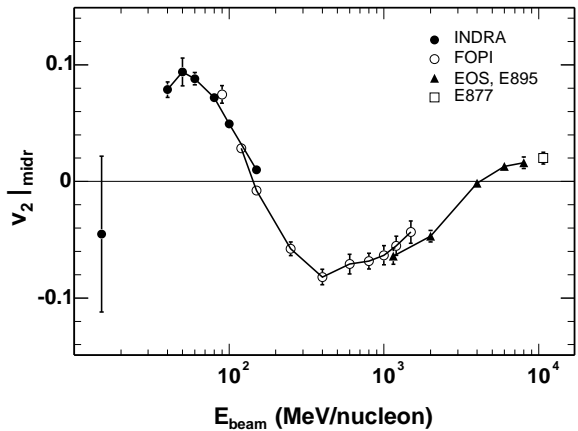}
 \end{center}
  
  \caption{Elliptic flow parameter $v_{2}$ at mid-rapidity for collisions at
  intermediate impact parameters (about 5.5-7.5 fm) as a function of incident
  energy, in the beam frame. The filled and open circles represent the INDRA
  and FOPI \protect\cite{andronic05} data, respectively, for $Z=1$ particles,
  the triangles represent the EOS and E895 \protect\cite{pinkenburg99} data for
  protons and the square represents the E877 data \protect\cite{bmunzinger98}
  for all charged particles.}

\label{fig:v2corr}
\end{figure}

The results on $v_2$ measured at midrapidity are summarized in fig.
\ref{fig:v2corr}. Elliptic flow varies as a function of energy from a
preferential in-plane, rotational-like \cite{tsang84, wilson90, lacey93},
emission ($v_2>0$) to an out-of-plane, or `squeeze-out' \cite{gutbrod90}
($v_2<0$) pattern, with a transition energy of about $150A$ MeV. This
transition energy is larger than that for the directed flow (see above and the
discussion in ref. \cite{crochet97}) and was shown to depend on centrality,
particle type and transverse momentum \cite{andronic01npa, andronic05}. For
higher energies, the strength of the collective expansion overcomes the
rotational-like motion, leading to an increase of out-of-plane emission. A
maximum is reached at $400A$ MeV, followed by a decrease  towards a transition
to preferential in-plane emission \cite{pinkenburg99, bmunzinger98}. This
behavior is the result of a complex interplay between fireball expansion and
spectator shadowing \cite{andronic05}, with the spectators acting as clocks of
the expansion times. For instance, in the energy range $400A-1500A$ MeV, the
passing time  of the spectators decreases from 30 to 16 fm/c, implying that
overall the expansion gets about two times faster in this energy range. This
interpretation is supported by the observed scaling of elliptic  flow as a
function of transverse momentum scaled with beam momentum \cite{andronic05}. We
note that the energy dependence of elliptic flow is similar to that of directed
flow \cite{reisdorf97, herrmann99, reisdorf04}, with the extra feature of the
transition to in-plane flow at $4A$ GeV \cite{pinkenburg99}. This high energy
transition has received particular interest as it is expected to provide a
sensitive probe of the EOS at high densities \cite{danielewicz98}. At SPS and
RHIC energies, the in-plane elliptic flow is determined by the pressure
gradient-driven expansion of the almond shaped isolated fireball
\cite{ollitrault92} and is currently under intensive experimental investigation
\cite{voloshin03, retiere04, lacey05}.

The agreement between the corrected INDRA and \linebreak[4]
FOPI data is good. The INDRA results have been corrected using the new
method, including the correction for the multi-hit losses (see sect.
\ref{sec:corrections}). According to IQMD
calculations, the reaction plane correction for the lowest FOPI
energy of $90A$ MeV appears to be somewhat overestimated.
On the other hand,
this may partially compensate for the lack of corrections due to unresolved
tracks which were not applied for $v_2$ in the FOPI case. 
Overall, the differences between the 
corrections is small enough, so that comparisons of uncorrected data sets are 
already meaningful. A good agreement was found to exist for the
%It is important to point out that a very good agreement was found between 
INDRA \cite{lukasik04, lukasik05}, FOPI \cite{andronic05} and Plastic
Ball \cite{gutbrod90} data in the reference frame of the directed flow and
without the correction for reaction plane resolution \cite{lukasik04,
andronic05}.

A remarkable feature of the $v_2$ observable is that it allows to show a
continuous evolution over a region covering completely different reaction
mechanisms, from those dominated by the mean field near the deep inelastic
domain, and the multifragmentation in the Fermi energy domain towards the
participant-spectator regime at relativistic energies.

\section{Correlation between stopping and flow}
\label{sec:correlation}

\begin{figure}[!htb]
%\centering{\mbox{\hspace{-1.5cm}
%\epsfig{file=$HOME/myfopi/pxdirvartl-wci.eps,width=7.5cm}}}
 \leavevmode
 \begin{center}
  \includegraphics[width=0.9\columnwidth]{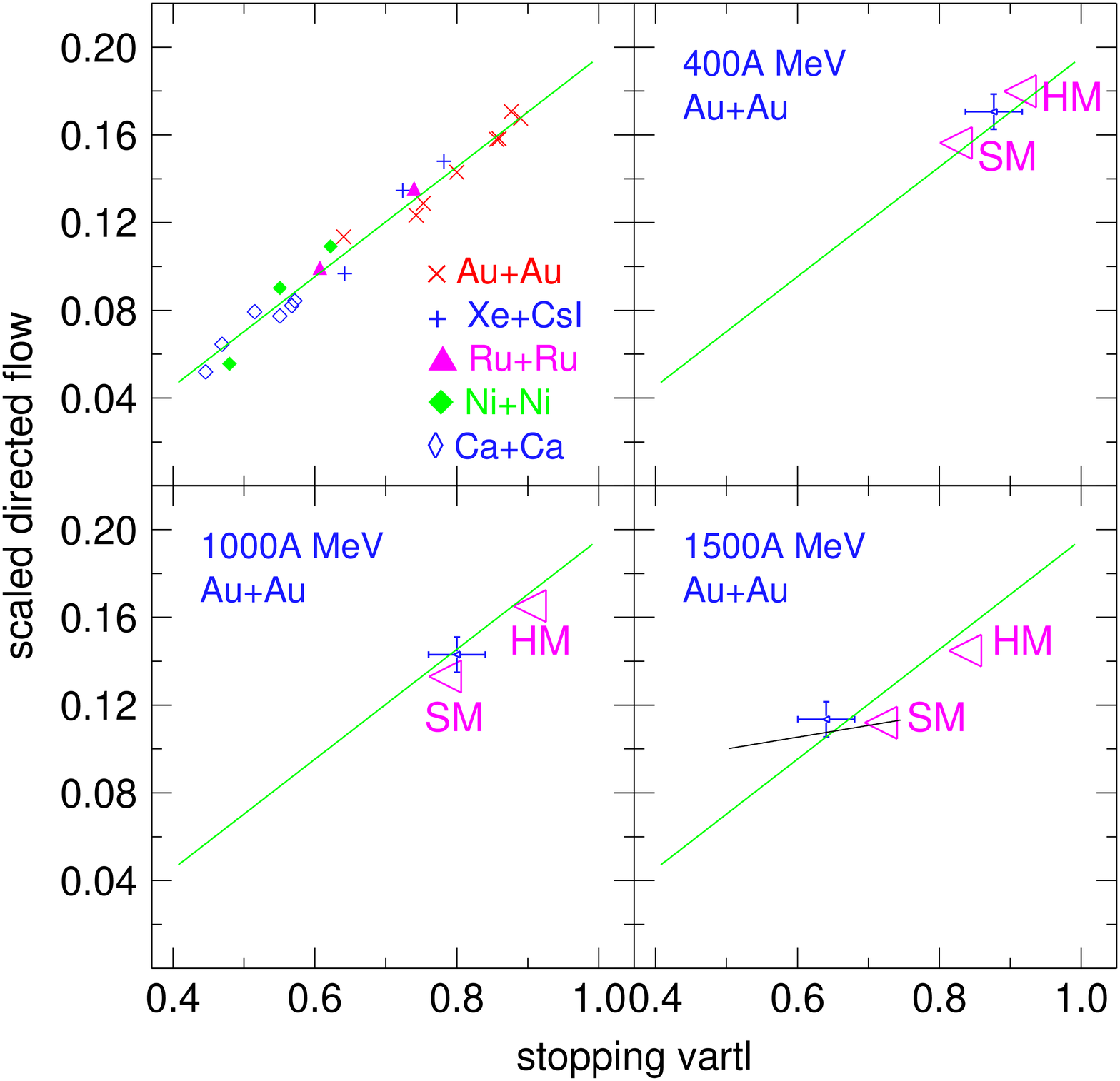}
 \end{center}
%\epsfig{file=pxdirvartl-wci.eps,width=10.cm}}}
%\vspace{-0.5cm}
%

\caption{Upper left panel: Correlation between the maximal directed sideflow
and the degree of stopping, after \protect\cite{reisdorf04}. The line is a
linear least squares fit to the data, which extend from $0.15A$ to $1.93A$ 
GeV. The correlation line is repeated in the other panels which show results of
simulations for Au+Au at three incident energies using two different equations
of state, $SM$ and $HM$, together with the experimental points. The short
segment passing through the $SM$ point in the lower right panel shows an
estimate of the trajectory using the $SM$ EoS and modifying the in medium cross
sections in a way that is compatible with \cite{gaitanos05}.}

\label{pxdirvartl}
\end{figure}

Information on stopping and flow in heavy ion collisions represents part of the
input to theoretical efforts to deduce constraints on the EoS. Remembering that
the EoS is a relation between pressure and density, it is intuitively
understandable that these two heavy ion observables are related to the EoS:
flow is generated by pressure gradients established in compressed matter, while
the achieved density is connected to the degree of stopping. Recently, it was
observed \cite{reisdorf04} that a strong correlation exists between the
stopping, measured in central collisions and the directed flow  measured at
impact parameters where it is maximal (see fig.~\ref{pxdir-prof}). The relevant
data are shown in fig.~\ref{pxdirvartl} in the upper left panel. Plotted
against each other are two dimensionless global event observables
characterizing stopping, $vartl$,  and global scaled directed flow,
$p_{xdir}^{(0)}$, both defined earlier.

\begin{figure}[!htb]
 \leavevmode
 \begin{center}
  \includegraphics[width=0.9\columnwidth]{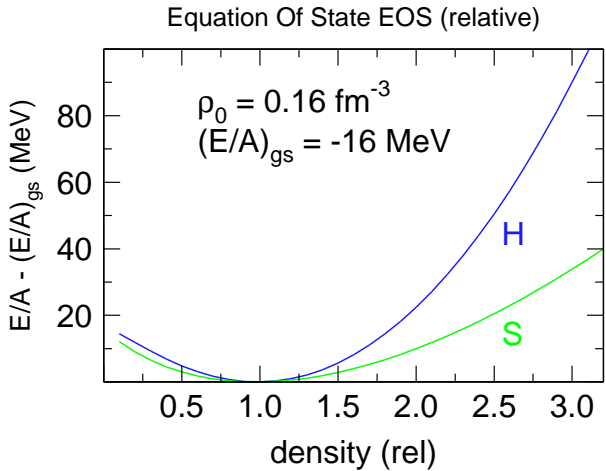}
 \end{center}

\caption{ Equations of state (relative to the ground state) used in the
calculation.}

\label{eos}
\end{figure}

The data points correspond to 21 system-energies with varying system size (from
Ca+Ca to Au+Au) and energy (from $150A$ to $1930A$ MeV). The straight
correlation line represents a linear least-squares fit to the data and is
repeated in the other panels. These other  panels show the location along the
correlation line of theoretical simulations using the IQMD code for Au + Au at
$400A$, $1000A$ and $1500A$ MeV as indicated. The points are marked $HM$ and
$SM$, respectively, for a stiff (incompressibility $K=380$ MeV) and a soft
($K=200$ MeV) EoS. The $M$ in $HM$ and $SM$ stands for the momentum dependence
of the $nn$ interaction. IQMD incorporates a phenomenological Ansatz fitted to
experimental data on the real part of the nucleon optical potential.  The
relevant experimental points are given together with their estimated systematic
errors (these errors were omitted for clarity in the upper left panel, but are
of comparable magnitude for all the data).  The main purpose is to show the
sensitivity of these combined observables to variations of the zero-temperature
EoS as compared to the uncertainty of the data. The EoS, that are purely
technical, are shown in fig.~\ref{eos}. The trajectory of the simulation when
changing $K$ seems to follow the correlation line, the distance between $SM$
and $HM$ is larger at $1000A$ MeV than at $400A$ MeV (i. e. the sensitivity is
increased at the higher energy), but then does not seem to further increase at
the highest energy, possibly due to the increase of transparency suggested by
fig.~\ref{vartl}.
Data measured at energies below $150A$ MeV do not continue the same linear
correlation, an interesting topic that deserves further studies.

In our exploratory IQMD simulations~\cite{hartnack98,leifels} we have not tried
to be realistic with regard to in-medium modifications of the nucleon nucleon
cross sections $\sigma_{nn}$, using instead the vacuum values standardly
implemented in the code~\cite{hartnack98}. An estimate of the trajectory in the
flow versus stopping plot when the EoS is kept constant, but the $\sigma_{nn}$
are decreased is shown in the right hand lower panel of fig.~\ref{pxdirvartl}.
For this estimate we rely on the more sophisticated calculations of
ref.~\cite{gaitanos05} which show that a switch to more realistic smaller
$\sigma_{nn}$ decreases the stopping by about $20\%$ and the scaled sideflow by
$6-7\%$, {\em i.e.} $\sigma_{nn}$ acts more strongly, relatively speaking, on
the stopping than on the scaled flow, as expected. The $\sigma_{nn}$
modification trajectory crosses the correlation line because it has a
different, flatter, slope than the EoS modification trajectory joining the $SM$
to the $HM$ point (which is not plotted) which happens to have a slope very
similar to that of the experimental correlation line. 
Generally speaking, one can say that
an underestimation of the apparent transparency will lead to an underestimation
of the stiffness of the EoS. Nevertheless, the procedure just outlined suggests
that an EoS closer to $SM$ than $HM$ would seem to be more appropriate to
describe the data. The same conclusion was reached from the comparison of the
FOPI data on directed flow, including its $p_T$ dependence, to IQMD
calculations \cite{andronic03} and from the comparison of the out-of-plane
expansion to BUU calculations \cite{stoicea04}.

Despite this encouraging result we would like to stress at this time that it
would be premature to draw firm conclusions from one particular transport code 
and it is  beyond the  scope of this experimental contribution to the  subject
to conclusively settle the question of the EoS.  Besides trying to predict
correctly the global observables just shown, probably a good strategy to start
with, the simulations must then proceed to reproduce the more differential data
such as the variations of the stopping and of 
the various flow components with the
particle type, as shown here in figs. \ref{vartlz} and \ref{fig:v1corr},
respectively. Another important physics quantity one would like to have under
theoretical control, in order to be convincing on the conclusion side, is the
created entropy. Although this is not a direct observable, the entropy at
freeze-out is strongly constrained by the degree of clusterization (of which we
showed an example in fig. \ref{dndz}) and the degree of pionization. An idea of
the freeze-out volume can be obtained from two-particle correlations
\cite{kotte05}, or even multi-particle correlations \cite{piantelli05,
tabacaru06}. All this is a rather challenging task. We refer to the work of
Danielewicz, Lacey and Lynch \cite{danielewicz02} for a summary of the
situation obtained a few years ago using a subset of the then available  heavy
ion data reaching up to the AGS energies.

%-------------------------------------------------------------------------

\section{Summary and outlook}
\label{sec:outlook}

We have presented a systematics of directed and elliptic flow and of stopping
for  $^{197}$Au + $^{197}$Au reactions in the intermediate range of energies
from  40 to 1500 MeV per nucleon by merging the data from INDRA and FOPI 
experiments performed at the SIS synchrotron at GSI. The overlap region of the
two data sets, 90 to 150 MeV per nucleon incident energy, has been  used to
confirm their accuracy on an absolute scale, and a very satisfactory  agreement
has been found.

Particular emphasis was given to the experimental reconstruction of the impact
parameter and to the corrections required by the dispersion of the
reconstructed azimuthal orientation of the reaction plane. The superiority of
observables based on transverse energy, either the ratio $Erat$ of  transverse
to longitudinal energy or the transverse energy $E_{\perp}^{12}$ of light
charged particles with $Z\leq 2$, over multiplicity for the selection of
central collisions has been demonstrated. A new method, derived by extending
the Gaussian approximation of the sub-$Q$-vector distributions to  the
non-isotropic case and by including the effect of correlation between the
sub-events, has been presented and applied to the data at the  lower incident
energies at which the multiplicities are still moderate  and the range of
emitted fragment $Z$ is still large, even in the most  central collisions. The
differences between the standard and the new corrections of derived flow
parameters are significant up to  incident energies as high as 150 MeV per
nucleon.

The deduced excitation functions of the $v_1$ and $v_2$ observables describing
directed and elliptic flow exhibit several changes of sign which reflect
qualitative changes of the underlying dynamics as a function of the  bombarding
energy. The transition from mean-field dominated attractive sidewards flow to
repulsive dynamics is observed for $Z=1$ and $Z=2$ particles at 80 MeV and 60
MeV per nucleon, respectively, in mid-central collisions. The transition from
predominantly in-plane to out-of-plane emissions occurs at 150 MeV per nucleon
for $Z=1$ particles. The second change-of-sign at several GeV per nucleon marks
the transition to the ultrarelativistic regime. These transition points are
quite well established and not very sensitive to  the chosen correction method.
The present study shows that also the maxima reached by the flow parameters are
reliable within the typically 5\%  systematic uncertainties due to the
corrections and the impact parameter selection. Within this margin they may be
used to test transport-model predictions and their sensitivity to the chosen
parameterization of the nuclear EoS.

It has, furthermore, been shown that the significance of the comparison can be
enhanced by including the experimentally observed stopping as represented by
the ratio of the variances of the integrated transverse and longitudinal 
rapidity distributions. This observable can best be determined for  central
collisions at which the directional flow vanishes for symmetry reasons whereas
the compression in the collision zone presumably reaches its maximum. The
common origin of the observed stopping and flow is evident from the strict
correlation of the two observables, including finite size effects. However,
their individual sensitivity to the magnitude of the nucleon-nucleon cross
sections and to the flow parameters is different and can be used to resolve
ambiguities between these two main ingredients of the models.  The sensitivity
to parameters of the equation of state is shown to increase with bombarding
energy over the present energy range, and a soft EoS  is clearly favored by the
data.

Further constraints for the determination of the parameters of the equation of
state can be obtained by  including the detailed dependences of flow on the
fragment $Z$, the impact parameter and the accepted ranges of transverse
momentum and rapidity into the comparison with theory. These data, for the
present reactions,  are either available already or in preparation. This will
have to be accompanied by theoretical studies of the still existing systematic 
differences between specific code realizations. The importance or necessity of
full antisymmetrization at low energies or of a covariant treatment  at high
bombarding energies and the role of nucleonic excitations will have to be
assessed.

On the experimental side, a gap of missing  flow data for the Au+Au system
exists at energies below 40 MeV per nucleon where interesting information on
transport coefficients as, {\em e.g.}, shear versus bulk viscosity or thermal
conductivity may be obtained. The origin of the observed negative flow should
be confirmed and clarified. At higher energies, new information, possibly also
on  the symmetry part of the  equation of state, can be expected from new
experiments involving  isotopically pure projectiles and targets and detector
systems permitting  mass identification at midrapidity.

\begin{acknowledgement} 
The authors would like to thank Y.~Leifels for the implementation of the  IQMD
code at GSI, the FOPI and INDRA-ALADIN Collaborations for the permission to
include partially unpublished data in this comparative study, and
J.-Y.~Ollitrault for stimulating discussions on flow evaluation and
corrections.

\end{acknowledgement}

\bibliographystyle{aipproc}   % if natbib is available

\begin{thebibliography}{999}
\bibitem{leifels} The code IQMD was implemented at GSI by Y. Leifels
(FOPI Collaboration), more results will be published elsewhere.
\end{thebibliography}

\end{document}